\documentclass[preprint,preprintnumbers,amsmath,amssymb]{revtex4}

\usepackage{graphicx}
\usepackage{dcolumn}
\usepackage{bm}
\usepackage{verbatim}
\usepackage{epsfig}
\usepackage{epstopdf}
\usepackage{amsmath}
\usepackage[caption=false]{subfig}	

\newcommand\beq{\begin{equation}}
\newcommand\eeq{\end{equation}}
\newcommand\bea{\begin{eqnarray}}
\newcommand\eea{\end{eqnarray}}

\usepackage{xcolor}

\def\x0{{{\bf x}_0}}

\begin{document}
	
	
	\title{Fine's theorem for Leggett-Garg tests with an arbitrary number of measurement times}

	\author{J.J.Halliwell}%
	\email{j.halliwell@imperial.ac.uk}
	\author{C. Mawby}
	\email{c.mawby18@imperial.ac.uk}
	\numberwithin{equation}{section}
	\renewcommand\thesection{\arabic{section}}

	\affiliation{Blackett Laboratory \\ Imperial College \\ London SW7
		2BZ \\ UK }

	
	
	\begin{abstract}
	If the time evolution of a quantum system can be understood classically, then there must exist an underlying probability distribution for the variables describing the system at a sequence of times.  It is well known that for systems described by a single time-evolving dichotomic variable $Q$ and for which a given set of temporal correlation functions  are specified, a necessary set of conditions for the existence of such a probability are provided by the Leggett-Garg (LG) inequalities.  Fine's theorem in this context is the non-trivial result that a suitably augmented set of LG inequalities are both necessary and sufficient conditions for the existence of an underlying probability.
We present a proof of Fine's theorem for the case of measurements on a dichotomic variable at an abitrary number of times, thereby generalizing the familiar proofs for three and four times. We demonstrate how the LG framework and Fine's theorem can be extended
to the case in which all possible two-time correlation functions are measured (instead of the partial set of two-time correlators normally studied). 
We examine the limit of a large number of measurements for both of the above cases.
			\end{abstract}


	
	\maketitle

	\section{Introduction}
	
There are a number of interesting situations in physics in which it is of interest to determine the conditions under which the ``partial snapshots'' of a given system, described by a set of marginal probabilities, are consistent with an underlying joint probability. Perhaps the most important and well-known examples of such conditions are the Bell and CHSH inequalities \cite{Bell,CHSH,Per} and their temporal analogue, the Leggett-Garg inequalities \cite{LeGa,L1,ELN}.  These conditions are of interest since they determine whether certain data sets can be simluated by classical stochastic models. Or equivalently, they establish the location of the boundary between classicality and quantumness. Such conditions are also of interest in studies of contextuality more broadly, both in physics \cite{ConPhys,AbBr} and beyond, e.g. in psychology \cite{ConPsy,DK}.

A particularly important result in this area is Fine's theorem \cite{Fine}, devised originally in the context of the CHSH inequalities, which establishes necessary and sufficient conditions for the existence of a joint probability matching a specific set of marginals.  A number of proofs of results of this sort of result have been given 
\cite{Bus,SuZa,Pit,GaMer,ZuBr,AbBr,JHFine}.
In this paper we derive generalizations of Fine's theorem for a number of situations of interest in the context of the Leggett-Garg (LG) inequalities.

The LG framework \cite{LeGa,L1,ELN} tests macroscopic realism (macrorealism), which is the view that a system evolving in time possess definite properties independent of past or future measurements. It is tested in particular systems by
measurements of a time-evolving dichotomic variable $Q$, taking values $s=\pm 1$,
in experiments at pairs of times chosen from the set $\{ t_1, t_2 \cdots t_n\}$. These measurements, which are assumed to be non-invasive, determine a particular set of pairwise probabilities $p(s_1, s_2)$, $p(s_2, s_3)$, \ldots, $p(s_{n-1}, s_n), p (s_1, s_n)$. (This is not the largest set of possibilities for pairwise measurements and we return to this point below.) 
These pairwise probabilities are required to be compatible with each other in the sense that, for example,
\beq
\sum_{s_1} p(s_1, s_2) = \sum_{s_3} p(s_2, s_3).
\label{cons}
\eeq
Each pairwise probability has a temporal correlation function of the form,
\beq
C_{12} = \sum_{s_1 s_2}  s_1 s_2 p(s_1, s_2),
\eeq
which may also be written $C_{12} = \langle Q_1 Q_2 \rangle$, where $Q_i$ denotes $Q(t_i)$.
Macrorealism implies that the set of pairwise probabilities possess an underlying joint probability $p(s_1,s_2 \cdots s_n)$, from which it readily follows that the correlation functions must obey a simple set of inequalities, similar in form to the Bell and CHSH inequalities for the cases $n=3$ and $n=4$. In particular,
for the case $n=3$ they are, 
   \bea
    C_{12} + C_{13} - C_{23} \le 1,
    \label{b1}
    \\
    C_{12} - C_{13} + C_{23} \le 1,
    \label{b2}
    \\
    - C_{12} + C_{13} + C_{23} \le 1,
    \label{b3}
    \\
    - C_{12} - C_{13} - C_{23} \le 1.
    \label{b4}
    \eea
These we will refer to as the three-time LG inequalities. (We will use the short-hand ${\rm LG}_3$ for the four three-time inequalities and similarly ${\rm LG}_n$ for the $n$-time case).
They are necesary conditions for macrorealism. Fine showed that inequalities of this type are also sufficient conditions for the existence of an underlying joint probability, as long as the pairwise marginals are non-negative and satisfy compatibility conditions of the form Eq.(\ref{cons}) \cite{Fine}.

The compatibility conditions of the form Eq.(\ref{cons}) deserve some attention in LG experiments. Whilst in Bell experiments their analogues (which refer to a pair of distant particles)  are expected to hold on the grounds of locality, in the LG framework they will not necessarily hold for sequential measurements and the conditions of Fine's theorem are then not met. One way around this is to use sequential measurements, which yield a non-negative probability, and then restrict the parameter space to values for which conditions of the form Eq.(\ref{cons}) are satisfied \cite{KoBr,Cle}. A second way \cite{HalQ,HalLG4} is to determine the pairwise probabilities using an indirect procedure in which, for measurements at times $t_1$ and $t_2$ for example, the correlator $C_{12}$ and the averages $\langle Q_1 \rangle$, $\langle Q_2 \rangle$ are determined in three separate non-invasive experiments and then the probability is assembled from the moment expansion, 
    \beq
    p(s_1, s_2)=\frac14 \left(1+ \langle Q_1 \rangle s_1 +  \langle Q_2 \rangle s_2 + C_{12} s_1 s_2\right).
\label{ps2}
    \eeq
(These convenient expansions are described in Refs.\cite{HaYe,Kly}.)
If all pairwise probabilities are determined in this way they will automatically satisfy the compatibility conditions of the form Eq.(\ref{cons}). This probability is not guaranteed to be non-negative (and indeed may be negative in quantum mechanics) so a set of conditions of the form $p(s_1, s_2) \ge 0 $ must also be imposed, which are often referred to as two-time inequalities (and are satisfied for a macrorealistic theory). 

These two different approaches to meeting the compatibility conditions of the form Eq.(\ref{cons}) actually correspond to different notions of macrorealism, as argued in Ref.\cite{HalLG4}, although both are clearly of interest to test. (Other variants of these two types of conditions exist \cite{WLG}.)
In the present paper we have in mind the weaker notion involving the two-time LG inequalities.
For the $n=3$ case this means that the necessary and sufficient conditions for the existence of a joint probability distribution consist of the four three-time LG inequalities adjoined with twelve two-time LG inequalities of the form $p(s_i, s_j) \ge 0 $, where $ij$ are the pairs $12$, $23$, $13$, with a similar statement for the four-time LG inequalities \cite{HalQ,HalLG4}.
The main aim of the present paper is to generalize this result to $n$-time measurements and establish the form of the LG inequalities in this case.

In Section 2, to set the stage, we give a streamlined version Fine's theorem for the case of pairs of measurements taken from measurements made at three or four times, based on previously given proofs \cite{Fine,JHFine} and in particular on Fine's ansatz.
In Section 3 we show how Fine's ansatz can be generalized to an arbitrary number of measurement times. We use this ansatz to prove Fine's theorem for LG inequalities at arbitrarily many times, using an inductive proof, and in the process, deduce the correct form of the complete set of LG inequalities in this case.

LG inequalities for $ n \ge 4$ always involve less than the complete set of two-time correlation functions (for example, in the familiar $n=4$ case there are a total of six correlators but only four appear in the LG inequalities). This rasises the question of necessary and sufficient conditions for the existence of an underlying probability when all possible two-time correlators are fixed. We prove some results for this case in Section 4, making contact with the ``pentagon inequality'' derived in Ref.\cite{Avis}.

In Section 5 we examine condition of both of the above types in the large $n$ limit, in which the LG inequalities appear to become easier to satisfy.
We summarize and conclude in Section 6.

    \section{Fine's Theorem for the LG Inequalities at Three and Four Times}

\subsection{Three-Time Case}
    
    In the three-time case, we are tasked with finding the conditions under which we can find a probability $p(s_1, s_2, s_3)$, which matches the three non-negative marginals $p(s_1, s_2)$, $p(s_2, s_3)$ and $p(s_1, s_3)$. Hence, this joint probability must be such that,       
    \beq
    p(s_1, s_2)=\sum_{s_3}p(s_1, s_2, s_3),
    \eeq
and likewise for $p(s_2,s_3)$ and $p(s_1,s_3)$.   We proceed using the moment expansion \cite{HaYe,Kly} of the three-time probability,
    \beq
    p(s_1, s_2, s_3) = \frac{1}{8}\left(1+\sum_i B_i s_i +\sum_{i<j}C_{ij}s_i s_j + Ds_1 s_2 s_3\right),
    \label{qs2}
    \eeq
    where $i,j = 1,2,3,$ and the coefficients are defined by 
    \bea
    B_i &=& \sum_{s_1 s_2 s_3}s_i  \ p(s_1, s_2, s_3) = \langle Q_i \rangle,
    \\
    C_{ij}&=&\sum_{s_1 s_2 s_3}s_i s_j \ p(s_1, s_2, s_3) = \langle Q_i Q_j \rangle,
    \\
    D &=&\sum_{s_1 s_2 s_3}s_1 s_2 s_3 \ p(s_1, s_2, s_3) = \langle Q_1 Q_2 Q_3 \rangle. \label{EqD}
    \eea
    It is readily seen from the moment expansions of the form Eq.(\ref{ps2}) that Eq.(\ref{qs2}) matches the three marginals.
    
    Since the coefficients $B_i$ and $C_{ij}$ are fixed, the question is whether or not the coefficient $D$ may be chosen so that the unifying probability Eq.(\ref{qs2}) is non-negative. We prove that a necessary and sufficient set of conditions for this are the three-time LG inequalities Eq.(\ref{b1})-(\ref{b4}).

    Necessity is easy to establish. To prove sufficiency note that Eq.(\ref{qs2}) is non-negative as long as,
    \beq
    A(s_1, s_2, s_3) \equiv 1 + \sum_i B_i s_i + \sum_{i < j} C_{ij} s_i s_j  \ge - D s_1 s_2 s_3.
    \label{A}
    \eeq
    For the four values of $s_1, s_2, s_3 $ for which $s_1 s_2 s_3 = - 1$, Eq.(\ref{A}) gives
    four upper bounds on $D$,
    \beq
    A(s_1, s_2, s_3 ) \ge D,
    \eeq
    and for the values with $s_1 s_2 s_3 = 1$, this give four lower bounds on $D$
    \beq
    D \ge - A(s_1, s_2, s_3). 
    \eeq
    Hence a value of $D$ exists as long as all four upper bounds are greater that the
    all four lower bounds. This yields sixteen inequalities which are readily shown \cite{JHFine} to be the four three-time LG inequalities Eqs.(\ref{b1})-(\ref{b4}),
    together with the twelve conditions $p(s_i,s_j) \ge 0 $ already assumed. 
   
   A natural question is whether the above upper and lower bounds on $D$ are compatible with the requirement $| D | \le 1 $, which follows from Eq.(\ref{EqD}).   It is readily seen that this is the case as long as $A(s_1,s_2,s_3) \ge - 1$. It is not immediately obvious that this relation holds, but this may be shown as follows. First, the conditions $p(s_i,s_j) \ge  0$ imply that
   \beq
   p(s_1,s_2) + p(s_2,s_3) + p(s_1,s_3) \ge 0,
   \eeq
   which may be written
   \beq
   3 +  2 \sum_i B_i s_i + \sum_{i < j} C_{ij} s_i s_j \ge 0. \label{3p}
   \eeq 
We also have the three-time LG inequalities, which may be written,
\beq
1 + \sum_{i<j} C_{ij} s_i s_j \ge 0. \label{LGx}.
\eeq    
Adding Eq.(\ref{3p}) and Eq.(\ref{LGx}), we obtain,
\beq
2 +  \sum_i B_i s_i + \sum_{i < j} C_{ij} s_i s_j \ge 0,
\eeq    
which is precisely the condition $A(s_1,s_2,s_3) \ge -1$. Hence we find compatibility with $|D|  \le 1$.
This completes the proof.

    
    \subsection{Four-Time Case}
    
    In the four-time case, the task is to find necessary and sufficient conditions for the existence of a joint probability $p(s_1,s_2,s_3,s_4)$ matching the four marginals $p(s_1,s_2)$, $p(s_2,s_3)$, $p(s_3,s_4)$, $p(s_1,s_4)$. As we will establish, these conditions are the eight four-time LG inequalities:
     \bea
   -2 & \le & C_{12} + C_{23} + C_{34} - C_{14} \le 2,
   \label{CHSH1}
   \\
   -2 & \le & C_{12} + C_{23} - C_{34} + C_{14} \le 2,
   \label{CHSH2}
   \\
   -2 & \le & C_{12} - C_{23} + C_{34} + C_{14} \le 2,
   \label{CHSH3}
   \\
   -2 & \le & - C_{12} + C_{23} + C_{34} + C_{14} \le 2.
   \label{CHSH4}
   \eea
Necessity is again easy to establish.  Only four of the possible six marginals are fixed in this problem.  This means that although the four $B_i$ are fixed, the two correlators $C_{13}$ and $C_{24}$ are not. This matching problem may be solved using Fine's insightful ansatz,
    \beq
    p(s_1, s_2, s_3,s_4)=\frac{p(s_1, s_2, s_3)\ p(s_1, s_3, s_4)}{p(s_1, s_3)},
    \label{fineansatz}
    \eeq
which breaks the problem down into demonstrating the non-negativity of two three-time probabilities and a two-time probability. It is readily shown, by summing out the appropriate pairs of variables (with a judicious choice of the order in which this is done), that this ansatz matches the four marginals of interest. 

The three-time probability $p(s_1, s_2, s_3)$ is non-negative as long as its three two-time marginals $p(s_1,s_2)$, $p(s_2,s_3)$ and $p(s_1,s_3)$ are non-negative and as long as the four ${\rm LG}_3$ inequalities hold. These  inequalities may be written in the convenient form,
\beq
-1 + \left| C_{12} + C_{23} \right| \le C_{13} \le 1 - \left| C_{12} - C_{23} \right|,
\label{Cineq1}
\eeq
which puts bounds on the unfixed quantity $C_{13}$.
Similarly, the three-time probability $p(s_1, s_3, s_4)$ is non-negative as long as its three two-time marginals $p(s_1,s_3)$, $p(s_3,s_4)$ and $p(s_1,s_4)$ are non-negative and as long as the corresponding four ${\rm LG}_3$ inequalities hold, which may be written,
\beq
-1 + \left| C_{14} + C_{34} \right| \le C_{13} \le 1 - \left| C_{14} - C_{34} \right|.
\label{Cineq2}
\eeq
Note that the marginal $p(s_1,s_3)$ appears both in the demoninator of the Fine ansatz and also as a marginal of both three-time probabilities, and since it is not fixed, its non-negativity must be imposed as another restriction on $C_{13}$, which has the form,
\beq
-1 + | B_1 + B_3 | \le C_{13} \le 1 - | B_1 - B_3 |.
\label{Cineq3}
\eeq

Eq.(\ref{Cineq1}) and Eq.(\ref{Cineq2}) together imply that a value of $C_{13}$ may be chosen as long as the two lower bounds are less than the two upper bounds, which is equivalent to,
\beq
 \left| C_{12} \pm C_{23} \right| +  \left| C_{14} \mp C_{34} \right|\le 2.
 \eeq
These two equations are in fact a concise rewriting of the eight ${\rm LG}_4$ inequalities, Eqs.(\ref{CHSH1})-(\ref{CHSH4}), the desired result.

However, we must also ensure that the upper and lower bounds in Eq.(\ref{Cineq3}) are compatible with those in     
Eq.(\ref{Cineq1}) and Eq.(\ref{Cineq2}) which is by no means obvious. Fortunately this is ensured by the fact that the four fixed marginals are non-negative, from which follow the inequalities
\bea
p(s_1,s_2) + p(-s_2,s_3) & \ge & 0,
\\
p(s_1,s_4) + p(s_3, -s_4) & \ge & 0.
\eea
Written out more explicitly these read,
\bea
2 + B_1 s_1 + B_3 s_3 &\ge& - C_{12} s_1 s_2 + C_{23} s_2 s_3,
\\
2+ B_1 s_1 + B_3 s_3 & \ge &- C_{14} s_1 s_4 + C_{34} s_3 s_4.
\eea
From this we see the compatibility of the bounds in Eq.(\ref{Cineq3}) with those in the other two relations,
Eq.(\ref{Cineq1}) and Eq.(\ref{Cineq2}).
This completes the proof.

    \section{Generalization to an Abitrary Number of Times}
    
    \subsection{Generalized Fine Ansatz}
    
   Given that the four-time LG inequalities ensure that a non-negative probability $p(s_1, s_2, s_3, s_4)$ may be found, we can ask about extending Fine's theorem to the case $n=5$.   We thus seek a joint probability $  p(s_1, s_2, s_3, s_4, s_5) $ matching the five pairwise probabilities $p(s_1,s_2)$, $p(s_2,s_3)$, $p(s_3,s_4)$, $p(s_4, s_5)$ and $p(s_1,s_5)$. We note that this may be solved using the generalized Fine ansatz,
    \beq
    p(s_1, s_2, s_3, s_4, s_5)=\frac { p (s_1, s_2, s_3 , s_4) \  p(s_1, s_4, s_5)} { p(s_1, s_4) }.
\label{fine5}
    \eeq
It is readily shown, by summing out triplets of values of the $s_i$ (where $i=1, 2, \cdots 5$) in a judiciously chosen order, that this ansatz matches the five fixed pairwise probabilities. The problem therefore reduces to the question of establishing the non-negativity of the four-, three- and two-time probabilities appearing in the ansatz, which will involve the four- and three-time LG inequalities and the non-negativity condition on $p(s_1,s_4)$.

We will not solve this problem explicitly, but note that it is suggestive of the Fine ansatz for the $n$-time case, which we postulate to be, 
\beq
    \label{fineansatzgen}
    p(s_1, \ldots, s_{n+1}) = \frac{p(s_1, \ldots, s_{n})\ p(s_1, s_{n}, s_{n+1})}{p(s_1, s_{n})}
    \eeq
    It is readily shown that this matches the $n$ pairwise marginals of interest, $p(s_i, s_j)$, where $(i,j)$ take the values $(1,2), (2,3), .... (n-1,n), (1,n)$. We will use this ansatz to given an inductive proof of Fine's theorem for $n$ times.

 We note in passing that through iterative application of the Fine ansatz, the $n$-time case may be reduced to a set of three-time problems, in terms of which the ansatz has the form, 
    \beq
    p(s_1, \ldots, s_{n+1})=p(s_1, s_2,s_3) \  \prod_{i=1}^{n-2}\frac{p(s_1, s_{i+2}, s_{i+3})}{p(s_1, s_{i+2})}.
    \eeq
This means that all $n$-time LG inequalities may be reduced to sets of three-time inequalities. A similar observation was noted in Ref.\cite{Avis}.

\subsection{The LG Inequalities for An Arbitrary Number of Times}
    
    Based on the three and four-time inequalities (and the five-time inequalities given in Ref.\cite{ELN}), we postulate that the $n$-time LG inequalities can be written as the $2^{n-1}$ relations, 
    \begin{equation}
    \label{lgansatz}
    a_1 C_{12}+a_2 C_{23}+ \ldots + a_{n-1}C_{n(n-1)}+ a_n C_{1 n} \leq n-2,
    \end{equation}
    where the coefficients $a_1, \ldots, a_{n}$ take values $\pm1$, and we constrain the product of all the coefficients $a_i$ to be negative,
    \begin{equation}
    \prod_{i=1}^n a_i = -1.
    \end{equation}
    This allows us to write one of the coefficients in terms of the others, for example, $a_n = -a_1 a_2 \ldots a_{n-1}$.  We see that the LG inequalities involve all possible sums of correlation functions with coefficients $\pm 1$ with an odd number of minus signs.  (Some specific higher order LG inequalities have been written down previously, e.g. Refs. \cite{ELN,deco1,LGn2}).

Eq.(\ref{lgansatz}) are readily seen to be necessary conditions for the existence of an underlying probability. The proof of this proceeds from the inequality
\beq
a_1 s_1 s_2 + a_2 s_2 s_3 + \cdots a_{n-1} s_{n-1} s_n + a_n s_n s_1 \le n-2,
\eeq
where the $s_i$ take values $\pm 1$, which can be established by choosing a fixed set of values of the $s_i$ (such as setting them all equal to $+1$) and the considering the effect of flipping their signs. Averaging this inequality with an underlying probability on $s_1, s_2, \cdots s_n$ then yields 
Eq.(\ref{lgansatz}). We now establish sufficiency.

\subsection{Inductive Proof}
     
     Following the method of Section 2, we now use the Fine ansatz Eq.(\ref{fineansatzgen}) and the $n$-time LG inequalties Eq.(\ref{lgansatz}) to show that the sufficient conditions for non-negativity of $p(s_1, \cdots s_{n+1})$ are the $(n+1)$-time LG inequalities.
     The probability $p(s_1, \cdots s_n)$ is non-negative as long as the LG inequalities Eq.(\ref{lgansatz}) are satisfied. These may be written,
    \beq
    A(a_n) + a_n C_{1n}\leq n-2,
    \label{eqn:shorthandlgn}
    \eeq
    where the function $A(a_n)$ is
    \beq
    A(a_n)= a_1 C_{12}+a_2 C_{23}+ \ldots + a_{n-1}C_{n(n-1)}.
    \eeq
    Noting that the argument $a_n$ takes values $\pm1$ and also that $a_1 \ldots a_{n-1}=-a_n$, we see that $A(\pm)$ are the sums of correlators with an odd/even number of minus signs.  We can now rewrite inequality Eq.(\ref{eqn:shorthandlgn}) as upper and a lower bound on $C_{1n}$,
    \beq
    -(n-2)+A(-)\leq C_{1n}\leq (n-2)-A(+).
\label{B1}
    \eeq
Similarly, the probability $p(s_1, s_{n}, s_{n+1})$ is non-negative if a set of three-time LG inequalities hold which if written in the general form Eq.(\ref{lgansatz}), are
    \beq
    b_1 C_{1n}+b_2 C_{n(n+1)}-b_1b_2 C_{1(n+1)}\leq1,
    \eeq 
    where $b_1, b_2$ take values $\pm 1$. This may 
be rewritten as
    \beq
    b_1 C_{1n} + B(b_1) \leq 1,
    \eeq    
    where
    \beq
    B(b_1) = b_2 C_{n(n+1)} - b_1 b_2 C_{1(n+1)}.
    \eeq
    Note that $B(\pm)$ are the sums of $C_{n(n+1)}$ and $C_{1(n+1)}$ with an odd/even number of minus signs.  This can be rearranged to give another upper and a lower bound on $C_{1n}$:
    \beq
    -1 +B(-)\leq C_{1n}\leq 1 - B(+).
\label{B2}
    \eeq
    A value of $C_{1n}$ obeying both set of bounds Eqs.(\ref{B1}), (\ref{B2}) may then be found as long as
\bea
B(-)+A(+)  & \leq(n-2)+1,
\\
B(+)+A(-)  & \leq(n-2)+1.
\eea
These relations may be rewritten,
\beq
A(a_n) + B( - a_n) \le (n+1) - 2
\label{res}
\eeq
It is not difficult to see that the left-hand side consists of all possible sums of correlators with an odd number of minus signs hence we have basically achieved a condition of the form Eq.(\ref{lgansatz}) with $n$ replaced with $(n+1)$, as required.
To be more explicit, Eq.(\ref{res}) reads,
\beq
a_1C_{12}+a_2 C_{23}+\ldots + a_{n-1} C_{n(n-1)}+b_2 C_{n(n+1)} + a_n b_2C_{1(n+1)}\leq (n+1)-2.
\eeq
This is a sum over $(n+1)$ correlators with $(n+1)$ independent coefficients taking values $\pm 1$ whose product is $ a_1 \cdots a_n b_2^2  = -1$. This is precisely of the form Eq.(\ref{lgansatz}).

Finally, as in the four-time case, we must also confirm that the restrictions on $C_{1n}$ are compatible with the non-negativity of $p(s_1,s_n)$. Using the moment expansion of the probability $p(s_1,s_n)$, its non-negativity gives us the following upper and lower bounds on $C_{1n}$,
    \beq
-1+\left|B_1+B_{n}\right|\leq C_{1n} \leq 1- \left|B_1 - B_{n}\right|,
    \label{eqn:positivepp}
    \eeq 
    and these must be compatible with the bounds Eqs.(\ref{B1}), (\ref{B2}).
    That is we require that we are always able to pick a $C_{1n}$ that satisfies the three sets of inequalities.  
    Since the measured pair probabilities are taken to be non-negative, we can add them and form new inequalities, 
    \beq
    p(s_1, -s_2)+p(s_2, -s_3)+\ldots+p(s_{n-1}, -s_n)\geq0,
    \eeq
    which using the moment expansion may be written explicitly as
    \beq
    (n-2)+1+B_1 s_1-B_n s_n - \sum_{i=1}^{n-1}C_{i(i+1)}s_i s_{i+1}\geq0.
    \label{eqn:positiveppn}
    \eeq
    For the case $s_1=-s_n$, the sum may be expressed as a sum of correlators with an odd amount of minus signs, e.g. a member of $A(+)$, and conversely for $s_1=s_n$, the sum may be expressed as a member of $A(-)$.  With this observation, it is simple to show that inequalities~(\ref{eqn:positivepp}) and Eq.(\ref{B1}) are indeed compatible.
   
    To ensure compatibility with Eq.(\ref{B2}), a similar argument may be made using the sum of pair probabilities,
    \beq
    p(s_1, -s_{n+1})+p(-s_n, s_{n+1})\geq 0,
    \eeq
    yielding
    \beq
    2+B_1s_1 - B_ns_n-C_{1(n+1)}s_1 s_{n+1}-C_{n(n+1)}s_n s_{n+1}\geq0.
    \eeq
    This inequality takes the same form as (\ref{eqn:positiveppn}), and the same arguments can be made to show that the inequalities~(\ref{eqn:positivepp}) and Eq.(\ref{B2}) are compatible.
This completes the inductive step of the proof.

We now observe that the three-time inequalities Eqs(\ref{b1})-(\ref{b4}) we proved earlier also match the form of (\ref{lgansatz}), and hence act as the base case for the inductive proof.  We have therefore proved the $n$-time generalisation of Fine's theorem, that for any $n\geq3$, the joint $n$-time probability distribution is guaranteed to exist, as long as all $2^{n-1}$ $n$-time LG inequalities Eq.(\ref{lgansatz}) are satisfied, together with the $4n$ two-time LG inequalities consisting of the non-negativity conditions on the fixed pairwise probabilities.

\section{Inequalities involving all of the two-time correlators}

An interesting feature of the LG inequalities is that they in general involve only a {\it subset} of all possible two-time correlators. For the three-time case all three correlators are measured and an underlying probability sought, but for the four time case only four out of the six possible correlators are measured. In general, the LG inequalities at $n$ times involve $n$ two-time correlators out of a total possible number of $n(n-1)/2$.
This choice of using only a subset of the total set of correlators arose because LG experiments were devised by way of analogy to Bell tests, in which one carries out pairs of measurements on a pair of particles, but not two measurements on the same particle, which means that certain correlators are not relevant experimentally. However this restriction is irrelevant in LG tests since all pairs of measurements are carried out on the same particle and furthermore, there is no obvious barrier experimentally to measuring the set of {\it all} two-time correlators. {This naturally raises the question as to the form of necessary and sufficient conditions for an underlying probability in the case in which the full set of $n(n-1)/2$ two-time correlators are measured, not just the $n$ correlators measured in standard LG tests. Since far more data is specified in this case than in the usual LG case, we expect that the necessary and sufficient condtions will be stronger than any set of LG inequalities. This case is rarely considered in the LG literature (and in fact Ref.\cite{Avis} is the only paper we are aware of that considers it). It involves some new features compared to the standard LG case which we will describe.}

\subsection{General Properties}

Avis et al \cite{Avis} consider the following condition for the case of measurements made at five possible times involving a sum of all ten possible correlators:
\beq
2 + \sum_{i<j} C_{ij} \ge 0,
\label{Av}
\eeq
where $i,j = 1, 2, \cdots 5$. 
{They argue that this condition is not a consequence of any LG inequalities and also that it may be violated by quantum mechanics. They refer to it as a ``pentagon inequality'' (out of acknowledgement for its geometric origins using the cut polytope).}

We now examine conditions of this type systematically. A general class of relations of the form Eq.(\ref{Av}) are readily derived by noting that the correlators may be written $C_{ij} = \langle Q_i Q_j \rangle$ and using the simple relation,
\beq
\left\langle \left( \sum_{i=1}^n s_i Q_i \right)^2 \right\rangle \ge  
\begin{cases}
1   &\mbox{if} {\ n \  {\rm odd} },  \\ 
0  & \mbox{if} {\ n \ { \rm  even} }.
\end{cases}
\eeq
This is readily seen to be true for a macrorealistic theory since all the $Q_i$ (and the $s_i$) are $\pm 1$, and for $n$ even, all the terms in the sum may cancel, but in the odd case, there must always be one left over.
This in turn may be written,
\beq
n + 2 \sum_{i<j} s_i s_j C_{ij} \ge 
\begin{cases}
1   &\mbox{if} {\ n \  {\rm odd} },  \\ 
0  & \mbox{if} {\ n \ { \rm  even} }.
\end{cases}
\label{neweq}
\eeq
{We will refer to these conditions as ``$n$-gon inequalities''}.
For $n=3$, these are in fact just the three-time LG inequalities:
\beq
1 + s_1 s_2 C_{12} + s_2 s_3 C_{23} + s_1 s_3 C_{13} \ge 0.
\label{L123}
\eeq
For $n=4$ and $n=5$, {the $n$-gon inequalities} are in fact the same,
\beq
2 + \sum_{i<j} s_i s_j C_{ij} \ge 0,
\label{Av2}
\eeq
which is a clear generalization of Eq.(\ref{Av}). However, {the $n$-gon inequalities} in the $n=4$ case can be written as an average of four sets of three-time LG inequalities, namely the inequalities Eq.(\ref{L123}), averaged with the three other sets obtained by choosing the time pairs from the triples $(t_1,t_2,t_4)$, $(t_2,t_3,t_4)$ and $(t_1,t_3,t_4)$. This is not possible in the $n=5$ case, as indicated in Ref.\cite{Avis} and as we see explicitly below.
Hence it is only for $n=5$ (and above) that {$n$-gon inequalities}
become stronger than the LG inequalities.

An interesting observation concerns whether the {$n$-gon inequalities} may continue to be satisfied in quantum mechanics or not. Replacing the variables $Q_i$ with their quantum operator counterparts $\hat Q_i$, we see that for all $n$ we have
\beq
\left\langle \left( \sum_{i=1}^n s_i \hat Q_i \right)^2 \right\rangle \ge  0.
\label{qc}
\eeq
These equalities therefore have the form
\beq
n + 2 \sum_{i<j} s_i s_j C_{ij} \ge  0,
\eeq
for all $n$, where here the quantum correlators are given by $C_{ij} = \frac{1}{2}\langle \hat Q_i \hat Q_j + \hat Q_j \hat Q_i \rangle$. This means that the {$n$-gon inequalities} Eq.(\ref{neweq}) are most interesting for $n$ odd where there is clear difference between the classical and quantum cases. For $n$ even, the classical and quantum versions coincide, at least for quantum correlators given by the above formula. (There may be a difference in the quantum case if the correlators are obtained differently, by so-called degeneracy-breaking measurements \cite{deg,PQS}).

For $n=3, $ Eq.(\ref{qc}) does in fact give the Tsirelson bound \cite{Tsi} for the three-time LG inequalities,
\beq
s_1 s_2 C_{12} + s_2 s_3 C_{23} + s_1 s_3 C_{13} \ge - \frac{3}{2}.
\eeq
Similarly for $n=5$ we have
\beq
\frac{5}{2} +  \sum_{i<j} s_i s_j C_{ij} \ge  0,
\eeq
in contrast to the classical version Eq.(\ref{Av2}).

\subsection{Sufficient Conditions: A Conjecture}

{We suppose that measurements are made (as described in the Introduction) to determine all possible pairwise probabilities of the form $p(s_i,s_j)$, where $i<j$ and $i,j$ run over $n$ values. We will suppose that they are all non-negative, i.e. all possible two-time LG inequalities are satisfied. The question is then to determine the necessary and sufficient conditions for which there exists a joint probability $p(s_1,s_2, \cdots s_n)$ which matches all the pairwise probabilities.}

{
The $n$-gon inequalities Eq.(\ref{neweq}) are clearly necessary conditions and likewise all possible $n$-time LG inequalities. Since the $n$-time LG inequalities are equivalent to sets of three-time inequalities a natural conjecture is as follows: A set of sufficient conditions for the existence of a joint probability matching all possible pairwise probabilities consists of the $n$-gon equalities Eq.(\ref{neweq}) together with all possible three-time LG inequalities of the form
\beq
1 + s_i s_j C_{ij} + s_i s_k C_{ik} + s_j s_k C_{jk} \ge 0,
\label{LGcomp}
\eeq
where $i<j<k$. A simple combinatorial analysis shows that
there are $2^{n-1}$ $n$-gon inequalities and $2n(n-1)(n-2)/3$ three-time inequalities.
(There are also $2n(n-1)$ two-time LG inequalities but these are assumed satisfied, as stated).}

We prove this conjecture for the (essentially trivial) case $n=4$ and the non-trivial case $n=5$. Our proof is restricted to the special ``symmetric'' case in which correlators involving an odd number of variables are zero.
This is not necessarily that restrictive for at least two reasons. One is that, as seen in Ref.\cite{JHFine}, the symmetric case is typically enough to establish the form that sufficient conditions should take -- no significant new conditions arise when going to the general case. Secondly, one can see from a quantum-mechanical analysis that the odd correlators can be made to vanish by choice of initial state. (This is accomplished by finding a reflection operator $R$ for which $ R \hat Q R = - \hat Q$, $ R H R = H$, where $H$ is the Hamiltonian, and chosing the state such that $R | \psi \rangle = | \psi \rangle $.)
This is not a significant restriction since the time spacings between measurements are the most important adjustable parameters in experimental tests.

\subsection{Sufficient Conditions: The Case $n=4$}

Consider then the case $n=4$. We seek a probability $p(s_1,s_2,s_3,s_4)$ matching all six correlators $C_{ij}$ where $ij = 12,13,14,23,24,34$. This is conveniently approached using the moment expansion,
\beq
p(s_1,s_2,s_3,s_4) = \frac{1}{16} \left(1 + \sum_{i<j} s_i s_j C_{ij} + E s_1 s_2 s_3 s_4 \right),
\eeq
for some constant $E$, where $-1 \le E \le 1$. This is non-negative as long as $E$ may be chosen so that
\beq
f(s_1,s_2,s_3,s_4) \equiv 1 + \sum_{i<j} s_i s_j C_{ij} \ge - E s_1 s_2 s_3 s_4.
\eeq
This reads
\beq
f(s_1,s_2,s_3,s_4) \ge  E,
\label{E1}
\eeq
for all values of the $s_i$ for which $ s_1 s_2 s_3 s_4 = - 1$ and reads
\beq
E \ge - f(s_1',s_2',s_3',s_4'),
\eeq
for all values for which $ s_1' s_2' s_3' s_4' = +1 $. Hence and $E$ may be found as long as all the lower bounds on it are less than all its upper bounds. This clearly leads to a set of conditions of the form,
\beq
f(s_1,s_2,s_3,s_4) + f(s_1,s_2,s_3,- s_4) \ge 0,
\eeq
plus three more with the minus sign in the other three places. (In general there would also be conditions with three sign flips but since $f(s_1,s_2,s_3,s_4)$ is symmetric under flipping all four signs this is equivalent to just one sign flip.) It is readily seen that these conditions are equivalent to the conditions that all four possible three-time probabilities $p(s_i,s_j,s_k)$  with $i<j<k$ are non-negative, which is guaranteed if all sets of three-time LG inequalities are satisfied, {as we have assumed}.
This proves the conjecture for $n=4$ in the symmetric case.

\subsection{Sufficient Conditions: The Case $n=5$}

For the $n=5$ case we seek a probability $p(s_1,s_2,s_3,s_4,s_5)$ matching all ten correlators $C_{ij}$. The moment expansion in the symmetric case is,
\beq
\begin{split}
p(s_1,s_2,s_3,s_4,s_5) = \frac{1}{32} \bigg(1 +& \sum_{i<j} s_i s_j C_{ij}  
+ E_1  s_2 s_3 s_4 s_5 
+ E_2 s_1 s_3 s_4 s_5
\\ 
+& E_3 s_1 s_2 s_4 s_5
+ E_4 s_1 s_2 s_3 s_5
+E_5 s_1 s_2 s_3 s_ 4
\bigg),
\end{split}
\eeq
for some constants $E_1, E_2, E_3, E_4, E_5$. We seek the conditions under which these constants may be chosen to ensure that the probability is non-negative. Following the method used in the $n=4$ case, we proceed to eliminate $E_1$ by
noting that the condition $p(s_1,s_2,s_3,s_4,s_5) \ge 0 $ may be written,
\beq
F(s_1,s_2,s_3,s_4,s_5) + E_1 s_2 s_3 s_4 s_5 \ge 0,
\eeq
where
\beq
\begin{split}
F(s_1,s_2,s_3,s_4,s_5) = 1 +& \sum_{i<j} s_i s_j C_{ij}  
+ E_2 s_1 s_3 s_4 s_5
+ E_3 s_1 s_2 s_4 s_5
\\
+& E_4 s_1 s_2 s_3 s_5
+E_5 s_1 s_2 s_3 s_ 4.
\end{split}
\eeq
In close analogy to the $n=4$ case, we readily find that the upper bounds on $E_1$ are greater than the lower bounds, and so a suitable value of $E_1$ may be chosen, as long as firstly,
\beq
F(s_1,s_2,s_3,s_4,s_5) + F(s_1,s_2,s_3,s_4,-s_5) \ge 0,
\label{F1}
\eeq
plus three more similar conditions in which $s_2$, $s_3$ or $s_4$ have their sign flipped; and secondly,
\beq
F(s_1,s_2,s_3,s_4,s_5) + F(s_1,s_2,-s_3,-s_4,-s_5) \ge 0,
\label{F2}
\eeq
plus three more conditions in which the triples $(s_2,s_3,s_4)$, $(s_2,s_4,s_5)$ or $(s_2,s_3,s_5)$ have all three signs flipped.

Eq.(\ref{F1}) has the form,
\beq
1 + \sum_{{i<j} \atop {i,j \ne 5}} s_i s_j C_{ij} + E_5 s_1 s_2 s_3 s_4  \ge 0,
\label{F3}
\eeq
which is precisely the $n=4$ case. It follows that
$E_5$ can be chosen to ensure that these inequalities are satisfied as long as a set of three-time LG inequalities holds. There will in addition be three more variants involving $E_2, E_3, E_4$. {It is readily shown that the three-time LG inequalities that need to be satisfied for Eq.(\ref{F3}) and its three variants to be satisfied is the complete set of three-time LG inequalities, Eq.(\ref{LGcomp}), with $ijk$ taking values $123, 124, 125, 134, 135, 145, 234, 235, 245, 345$.}

However, the upper and lower bounds on $E_5$ implied by Eq.(\ref{F3}) must be also compatible with the further upper and lower bounds on $E_5$ derived below, and likewise for the three variants of this inequality.

Similarly, Eq.(\ref{F2}) has the form,
\beq
\begin{split}
1 &+ s_1 s_2  C_{12} + s_3 s_4 C_{34} + s_3 s_5 C_{35} + s_4 s_5 C_{45}
\\
&+ E_3 s_1 s_2 s_4 s_5 + E_4 s_1 s_2 s_3 s_5 + E_5 s_1 s_2 s_3 s_4 \ge 0.
\end{split}
\label{F4}
\eeq
There will be three similar relationships for the three variants of Eq.(\ref{F2}), but will clearly be qualitatively the same by relabelling. We can solve this inequality by eliminating, say, $E_4$, by writing,
\beq
G(s_1,s_2,s_3,s_4,s_5) + E_4 s_1 s_2 s_3 s_5 \ge 0
\eeq
where
\beq
\begin{split}
G(s_1,s_2,s_3,s_4,s_5) =
1 &+ s_1 s_2  C_{12} + s_3 s_4 C_{34} + s_3 s_5 C_{35} + s_4 s_5 C_{45}
\\
&+ E_3 s_1 s_2 s_4 s_5 + E_5 s_1 s_2 s_3 s_4 
\end{split}
\label{G2}
\eeq
Proceeding as above, we readily find that a value of $E_4$ may be chosen as long as the following two sets of conditions hold. Firstly,
\beq
G(s_1,s_2,s_3,s_4,s_5) + G(s_1,s_2,s_3,s_4,-s_5) \ge 0,
\label{G3}
\eeq
plus three more similar conditions with $s_1$, $s_2$ or $s_3$ having their sign flpped; and second,
\beq
G(s_1,s_2,s_3,s_4,s_5) + G(-s_1,-s_2,-s_3,s_4,s_5) \ge 0,
\label{G4}
\eeq
plus three more conditions with the triples $(s_1,s_2,s_5)$, $(s_1,s_3,s_5)$ and $(s_2,s_3,s_5)$ having all three signs flipped.

Eq.(\ref{G3}) has the form,
\beq
1 + s_1 s_2 C_{12} + s_3 s_4 C_{34} + E_5 s_1 s_2 s_3 s_4 \ge 0.
\label{G5}
\eeq
It is easily shown that a value of $E_5$ may be chosen so that these inequalities are satisfied, relying only on the property $ |C_{ij} | \le 1 $ of the correlators. We must also confirm that Eq.(\ref{G5}) is compatible with other bounds on $E_5$ and in particular the bounds implied by Eq.(\ref{F3}). It is readily seen that this is the case subject only to a set of three-time LG inequalities. Similar statements apply to the three variants of Eq.(\ref{G3}).

Consider now the four inequalities of the form Eq.(\ref{G4}). Two of them are,
\bea
1 + s_1 s_2 C_{12} + s_3 s_4 C_{34} + E_5 s_1 s_2 s_3 s_4 &\ge& 0,
\\
1 + s_1 s_2 C_{12} + s_4 s_5 C_{45} + E_3 s_1 s_2 s_4 s_5 &\ge& 0,
\eea
which are straighforwardly handled as above. The other two inequalities turn out in fact to be the same and have the form
\beq
1 + s_3 s_5 C_{35} + E_3 s_ 1 s_2 s_4 s_5 + E_5 s_1 s_2 s_3 s_4 \ge 0.
\label{G8}
\eeq
This set of inequalities taken on its own can be satisfied for some $E_3$ and $E_5$ using $|C_{35}| \ge 1$.
However, the bounds on $E_3$ and $E_5$ implied by Eq.(\ref{G8}) must be compared with all other inequalities involving $E_3$ and $E_5$. A potentially lengthy search through all the possibilities is required to check all the cases, but it is not difficult to see that 
the only ones that give non-trivial new conditions are Eq.(\ref{F3}) and its variant involving $E_3$. These two inequalities may be written,
\bea
E_5 s_1 s_2 s_3 s_4 & \ge& - 1 - \sum_{{i<j} \atop {i,j \ne 5}} s_i s_j C_{ij},
\\
E_3 s_1 s_2 s_4 s_5  & \ge & - 1 - \sum_{{i<j} \atop {i,j \ne 3}} s_i s_j C_{ij}.
\eea
We compare these two inequalities to Eq.(\ref{G8}) with the signs of $s_3$ and $s_5$ reversed, which reads,
\beq
1 + s_3 s_5 C_{35} \ge  E_3 s_ 1 s_2 s_4 s_5 + E_5 s_1 s_2 s_3 s_4.
\eeq
The last three inequalities are compatible as long as,
\beq
1 + s_3 s_5 C_{35} \ge -2 - \sum_{{i<j} \atop {i,j \ne 3}} s_i s_j C_{ij}
- \sum_{{i<j} \atop {i,j \ne 5}} s_i s_j C_{ij}.
\eeq
Written out in full this inequality is conveniently written
\beq
\left( 1 + s_1 s_2 C_{12} + s_1 s_4 C_{14} + s_2 s_4 C_{24} \right) + 2 + \sum_{i<j} s_i s_j C_{ij} \ge 0,
\eeq
where the sum is over all $ij$. Noting that the first term in parantheses is a LG inequality so is non-negative, the total inequality is satisfied as long as
\beq
2 + \sum_{i<j} s_i s_j C_{ij} \ge 0.
\eeq
This is the desired result, Eq.(\ref{Av2}), and establishes the conjecture for $n=5$ in the symmetric case.

\subsection{Other Approaches}

Although as argued the symmetric case is not as restrictive as it might seem, it would clearly be desirable to extend the proof to the general case. Here we have focused on a purely algebraic proof but a geometric one involving the cut polytope geometry discussed in Ref.\cite{Avis} is clearly a natural place to look and may get a
away from the restriction to the symmetric case considered here. 

Another very different approach would be to do numerical experiments which confirm the conjecture.
We start with the moment expansion in the general case:
\beq
\begin{split}
p(s_1, s_2, s_3, s_4, s_5) = \frac {1} {32} \bigg( 1 +&
\sum_i B_i s_i + \sum_{i<j} C_{ij} s_i s_j + \sum_{i<j<k} D_{ijk} s_i s_j s_k
\\
+& \sum_{i<j<k<\ell} E_{i j k \ell}  s_i s_j s_k s_\ell 
+ F s_1 s_2 s_3 s_4 s_5 \bigg).
\end{split}
\label{qs}
\eeq
We then suppose that values of the fixed quantities $B_i$, $C_{ij}$ are chosen at random, which may or may not satisfy the requisite conditions (i.e. the full set of two and three-time LG inequalities and the generalized pentagon inequalities Eq.(\ref{Av2})). For each set of values of the fixed quantities, a numerical search is carried out to find
values of the coefficients $D_{ijk}$, $E_{ijk\ell}$ and $F$ for which the $32$ inequalities $p(s_1, s_2, s_3, s_4, s_5) \ge 0 $ are satisfied. 
The conjecture is upheld if the success or failure of the search is found to be in direct correspondence to whether the fixed values do or do not satisfy the requisite conditions.
{We have carried out a series of such numerical experiments using Mathematica and found no counter-examples to the conjecture. However, we find that the interpretation of such numerical experiments is subtle and it is difficult to see how to use them to make reliable claims. This will be addressed in more detail in future publications.}

\subsection{Further Generalizations}

A further natural generalization is to consider situations in which third and higher-order correlators are measured. The appropriate necessary and sufficient conditions that these must satisfy have been given in Refs.\cite{HalNIM,PQS}. These conditions are of experimental interest because
measurement of some of these higher-order correlators has been carried out \cite{Bec}.

\section{The Limit of a Large Number of Measurements}
We now make some observations about the simplifications and general features that arises for large $n$, for the ${\rm LG}_n$ inequalities Eq.(\ref{lgansatz}), and the $n$-gon inequalities Eq.(\ref{neweq}).

\subsection{Algebraic Argument}
If we assume that the $C_{ij}$ depend solely on the time difference $t_j - t_i$, and also make the common assumption of equal time spacing $\tau$ between the times $t_i$,
then we have $C_{12}=C_{23}=\ldots=C_{n(n+1)}$.  The general $n$-time LG inequalities may then be written as
\beq
C_\tau\left(a_1+\ldots+a_{n-1}\right) \pm C_{n\tau} \leq n-2,
\eeq
Now note that the sum of the $a_i$ coefficents takes extremal values of $\pm(n-1)$, so we may write the inequality in both worst cases as
\beq
-\frac{n-2}{n-1}\leq C_\tau + \frac{C_{n\tau}}{n-1}\leq \frac{n-2}{n-1}
\eeq
In the limiting case of large $n$, this inequality tends toward
\beq
-1\leq C_{\tau}\leq 1
\eeq
which is a condition the correlators satisfy already. Hence the ${\rm LG}_n$ become easier to satisfy for large $n$, becoming identically satisfied as $n$ goes to infinity. 
Mathematically, there is an intuitively clear reason why it might become easier to satisfy ${\rm LG}_n$ for large $n$. This is that the LG inequalities involve just $n$ correlators, but the probabilities for $n$ times have $2^n-1$ independent components so the requirement to match $n$ correlators becomes less restrictive, relatively speaking, as $n$ increases. However, this is perhaps counter to physical intuition, which suggests that it should be harder to assign probabilities to histories in which more variables are specified, especially in quantum mechanics.

There are two cases to consider. One is the case in which we have a fixed time interval $[0,T]$ and just add progressively more measurements in that interval. However, in that case, $C_{12}$ will tend to $1$ for large $n$ (since $t_2-t_1$ goes to zero) and then ${\rm LG}_n$ can only be satisfied for $C_{1n}$ = 1. 

The second case is that which we keep adding extra measurements at later times, so prolonging the total time interval, and in that case it does indeed seem that, as indicated, ${\rm LG}_n$ gets easier to satisfy as $n$ increases.
In LG tests we are trying to determine whether or not there exists a probability $p(s_1, s_2 \cdots s_n)$ corresponding to certain sets of measurements taken at $n$ times. In general, 
we would expect it to be harder to find a probability for $(n+1)$ times,  $p(s_1, s_2 \cdots s_n, s_{n+1})$, since that would be a fine graining, and easier to find a probability for $(n-1)$ times, $p(s_1, s_2, \cdots s_{n-1})$ since that is a coarse-graining.

However, LG tests refer to a very specific set of measurements in which only certain correlators are measured and as a consequence the probabilities it seeks for $(n-1)$, $n$ and $(n+1)$ times are not simply related. The $n$-time LG inequalities refer to the correlators $C_{12} \cdots C_{(n-1)n}$ and $ C_{1n}$ but the $(n+1)$-time inequalities refer to the correlators $C_{12} \cdots C_{n(n+1)}$ and $ C_{1(n+1)}$, which does not involve $C_{1n}$ but does involve two new correlators that do not appear in the $n$-time case. This means that the probabilities for $n$ times and $(n+1)$ times that these LG inequalities test for are not simple fine or coarse grainings of each other -- the fixed quantities in the $n$-time case are not a subset of the fixed quantities of the $(n+1)$-time case.

Clearly a case in which the set of fixed quantities {\it are} simply related subsets for different values of $n$ 
is that in which all possible two-time correlators are fixed, discussed in the previous section and one would therefore expect that the $n$-time probabilities would behave according to intuition under fine or coarse graining.
There, as demonstrated, we require not just the LG inequalities to hold but also the conditions 
Eq.(\ref{neweq}). Since the total number of two-time correlators is  $ n (n-1)/2$, Eq.(\ref{neweq}) will become {\it harder} to satisfy as $n$ increases, even if the LG inequalities become easier. This is the intuitively expected result.

\subsection{Approximate Measure of Violation from the Central Limit Theorem}

By looking at the case where measurements are spaced equally in time, we have provided a simple argument for the behaviour of the $n$-time LG inequalities in the large $n$ regime.  We now make a more sophisticated and general argument, which includes the case of arbitrarily spaced measurements.  This argument is applicable to both the ${\rm LG}_n$ and $n$-gon inequalities Eqs.~(\ref{lgansatz}), (\ref{neweq}), and more generally to any inequality formed as a sum of correlators.

We aim to establish for a given $n$, a measure of how easy, or hard it is to violate a given inequality from the family of ${\rm LG}_n$ or $n$-gon inequalities.  One approach would be to ask what volume of the parameter space of $C_{12},\ldots,C_{1n}$ leads to a violation of the given inequality. Owing to the equivalence between volume in parameter space, and the probability of a certain event happening, this question is profitably reframed as a question of probabilities.  If we pick random values for $C_{12},\ldots,C_{1n}$ from a uniform distribution, then the probability of that point leading to a violation of the inequality will exactly correspond to the volume in parameter space capable of supporting a violation.  Hence calculating this probability will give us the desired measure of the ease or difficulty of violation.

\begin{figure}
	\begin{center}
		\includegraphics[width=0.65\textwidth]{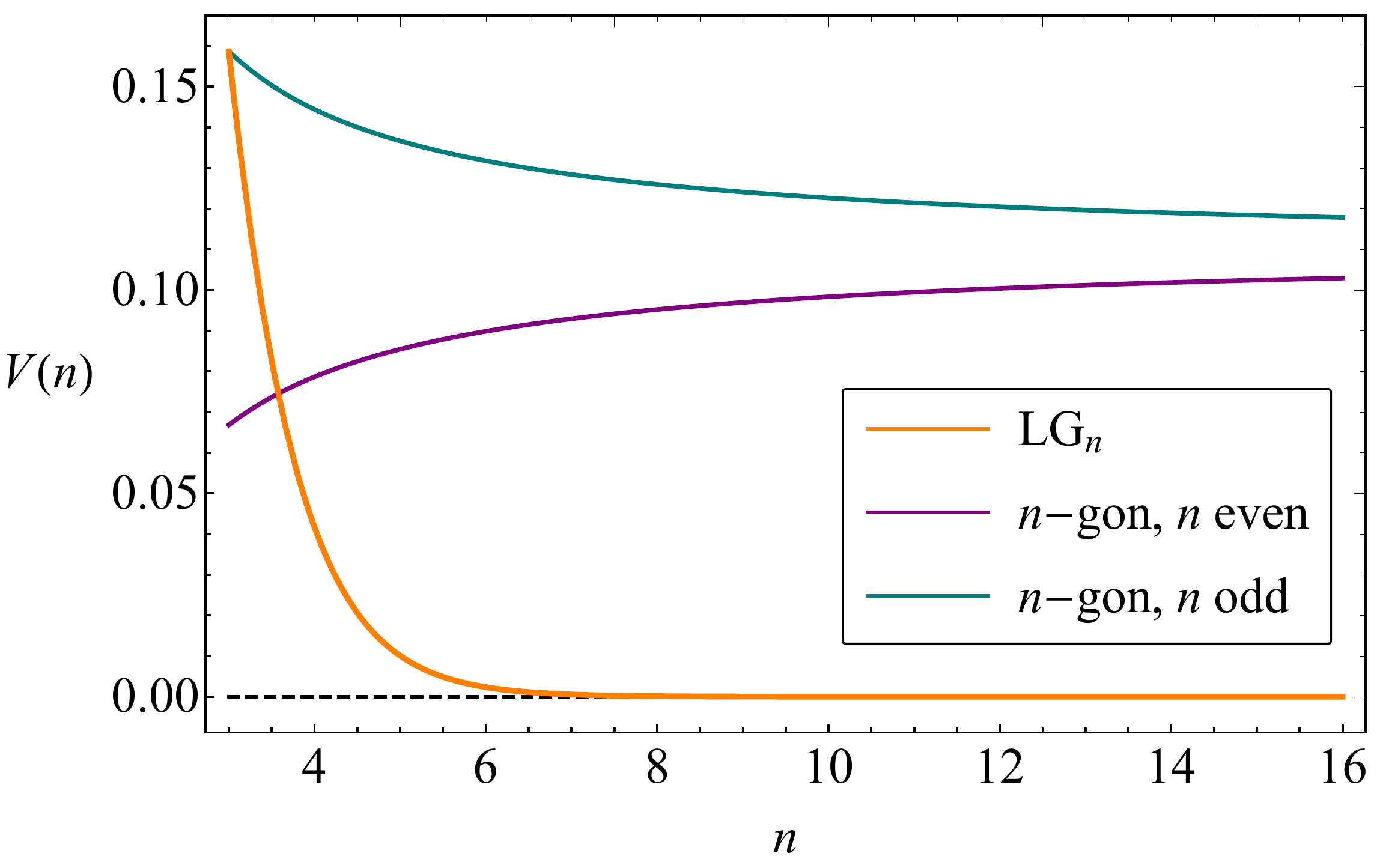}
	\end{center}
	\caption{
		\label{fig:clt}
		The fraction of parameter space which leads to the violation of a given ${\rm LG}_n$ or $n$-gon inequality, as estimated by the central limit theorem, Eq.(\ref{eq:vlgn}), Eq.(\ref{eq:vngon}).}
\end{figure}

Under this formulation of the problem, we treat the correlators as uniformly distributed random variables, taking values from $-1$ to $1$.  We now note that since the distribution is uniform, the distributions of $C_{ij}$ and $-C_{ij}$ are identical.  Hence when we consider the contribution from a given correlator, we see that probabilistically speaking, it does not matter whether it appears with a plus or a minus sign.  In the inequalities Eq.(\ref{lgansatz}) and Eq.(\ref{neweq}) where we see sums of correlators with different signs, we may instead consider just a single sum of correlators with all pluses.  Within the probabilistic formulation, the analysis of this one case corresponds to exploring the set of all possible sign permutations of correlators in the original inequality.  Then since the specific choice of signs appearing in the LG inequalities, or $n$-gon inequalities, will be contained within this set, we see the analysis of this one case is sufficient. The inequalities in which we are interested may all be analysed using the general form
\beq
\label{eq:probineq}
S_{j(n)}\leq b(n),
\eeq
where $S_j$ represents the sum of the $j$ correlators, represented by random variables, $j(n)$ is the number of correlators involved in the $n$-th order inequality, and $b(n)$ is the upper bound on the inequality.  We now invoke the central limit theorem (CLT) \cite{CLT} for a uniform distribution, which states that the distribution of $\frac{S_j\sqrt{3}}{\sqrt{j}}$ for the sum of $j$ independent uniformly distributed variables from $-1$ to $1$, will tend towards the standard normal distribution.  Comparing this result to the generalised inequality Eq.(\ref{eq:probineq}), we can estimate the region of violation as
\beq
V(n)=\frac{1}{\sqrt{2\pi}}\int_{\frac{b(n)\sqrt{3}}{ \sqrt{j(n)}}}^{\infty}e^{-\frac{x^2}{2}}\mathop{dx}.
\eeq
This integral is simply evaluated as
\beq
\label{eq:violation}
V(n)=\frac{1}{2}\left(1-\text{erf}\left(\sqrt{\frac{3}{2}}\frac{b(n)}{\sqrt{j(n)}}\right)\right).
\eeq
This result says that the difficulty or ease of violating any $n$-th order inequality formed of the sum of $j(n)$ correlators (with arbitrary signs on each correlator), with an upper bound $b(n)$, is determined solely by the functional form of the ratio of $\frac{b(n)}{\sqrt{(j(n))}}$.

Comparing Eq.(\ref{eq:probineq}) to the general LG inequality Eq.(\ref{lgansatz}), we see that for the LG case, we have $b(n)=n-2$, and $j(n)=n$.  Using these values in Eq.(\ref{eq:violation}), we find
\beq
\label{eq:vlgn}
V_{\text{LG}}(n)=\frac{1}{2}\left(1-\text{erf}\left(\sqrt{\frac32}\frac{n-2}{\sqrt{n}}\right)\right).
\eeq
For large $n$, the argument of the error function behaves as $\sqrt{n}$, and hence the violating region of parameter space vanishes with increasing $n$.  This behaviour is shown in Fig.~\ref{fig:clt}.

Comparing the form of the $n$-gon inequality Eq.(\ref{neweq}) to Eq.(\ref{eq:probineq}), we have $b(n)=\frac{n}{2}$ for $n$ even, $b(n)=\frac{n-1}{2}$ for $n$ odd, and $j(n)=\frac12n(n-1)$. In the case of even $n$, this leads to a violation region of
\beq
\label{eq:vngon}
V_{n-\text{gon}}(n)=\frac{1}{2}\left(1-\text{erf}\left(\frac{\sqrt{3}}{2}\frac{n}{\sqrt{n(n-1)}}\right)\right)
\eeq
In both the even and odd cases, for large $n$, the argument of the error function tends toward $\frac{\sqrt{3}}{2}$, and hence the violation asymptotes to $V\approx0.11$, as depicted in Fig.~\ref{fig:clt}.   


This CLT approximation makes precise the earlier arguments that ${\rm LG}_n$ inequalities become easier to satisfy for large $n$, and illuminates that the generalised $n$-gon inequalities maintain significant regions of violation for all $n$.



\subsection{Illustration in a Simple Spin Model}
\begin{figure}
	\subfloat[]{{\includegraphics[height=5.3cm]{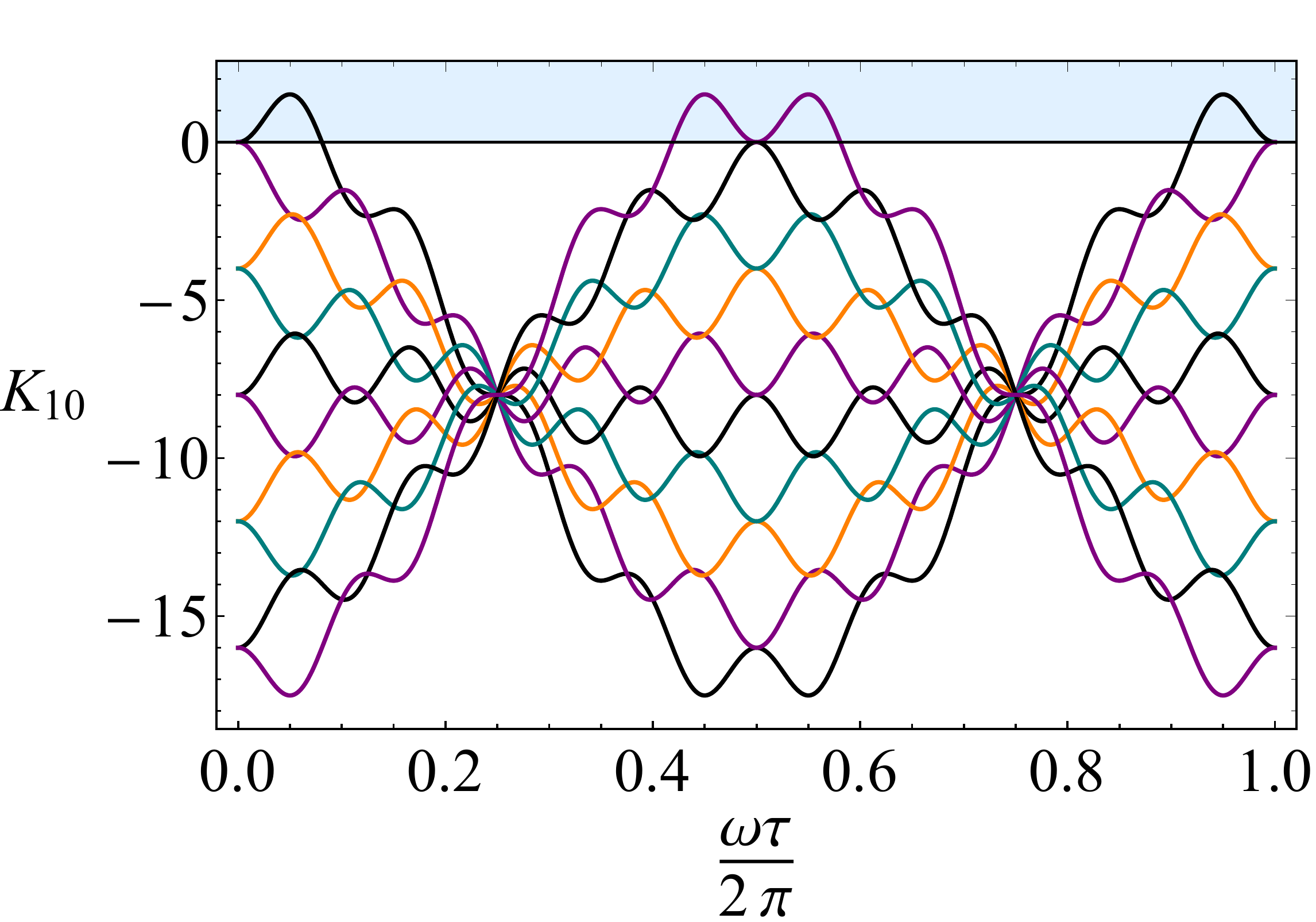}}}%
	\qquad
	\subfloat[]{{\includegraphics[height=5.05cm]{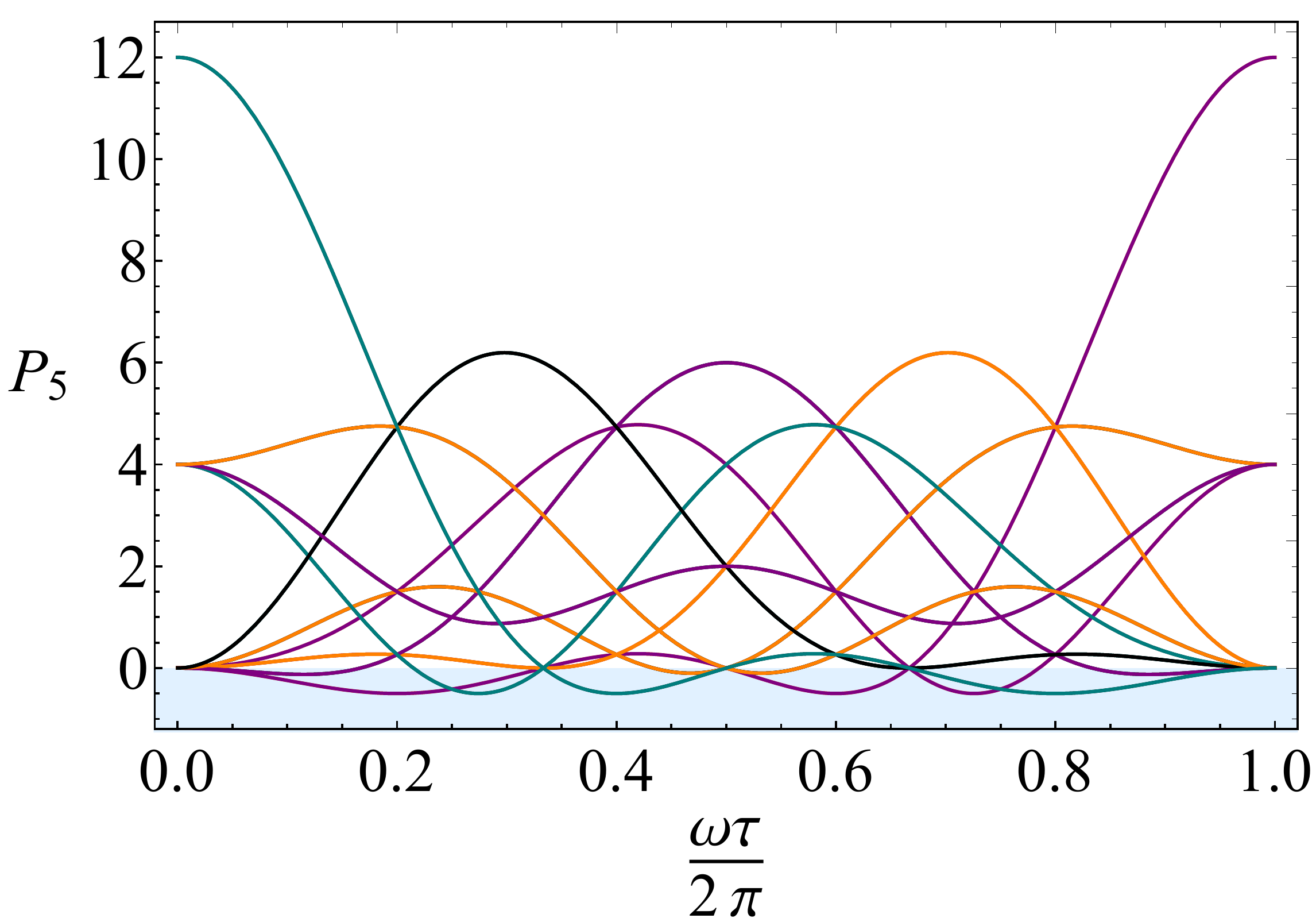}}}
	\caption{In (a), the 10 distinct inequalities from the $n=10$ LG case are plotted, for the simple spin system.  The inequalities are violated in the shaded region,  $K_{10}\geq0$.  In (b), the 10 distinct inequalities from the $n=5$ $n$-gon case are plotted.  In this case, at least one of the inequalities is violated for {each} $\tau$, except at points of measure zero.}%
	\label{fig:simplespin}%
\end{figure}
We now illustrate the above results using plots in the commonly-studied simple  spin model \cite{Ema}
in which the correlators are given by
\beq
\label{eq:corr}
C_{ij}=\cos\left(\omega(t_j-t_i)\right).
\eeq
We also make the common assumption of equal time spacing $\tau$.
Purely for convenience of plotting, we rearrange the LG inequalities, introducing the notation $K_n$, defined as
\beq
\label{eq:kernel}
K_n =  a_1 C_{12}+a_2 C_{23}+ \ldots + a_{n-1}C_{n(n-1)}+ a_n C_{1 n} - (n-2),
\eeq
and so violations of the ${\rm LG}_n$ inequalities Eq.(\ref{lgansatz}) are delineated by $K_n\geq0$.  We introduce a similar notation $P_n$ for the $n$-gon inequalities Eq.(\ref{neweq}), where $P_n\leq0$ signifies a violation.

Recall that in Eq.(\ref{eq:kernel}) $a_1, \ldots, a_n$ take all values of $\pm1$ such that their overall product is $-1$, and so we must study the behaviour of each member of the family of inequalities for a given $n$.  For example, for the $n=10$ case of the LG inequalities, there are a total of 512 inequalities.  When working with equally spaced measurements, it turns out that only 10 of these inequalities are distinct.  
These 10 inequalities are plotted for the simple spin case in Fig.~\ref{fig:simplespin}(a).  This plot can be considered representative behaviour of the ${\rm{LG}}_n$ inequalities in the simple spin case, where the bulk of the inequalities are always satisfied.  Furthermore, the only two that are not satisfied are violated over only a small region of measurement times.

For the $n$-gon case for $n=10$ there are 1024 inequalities, of which 272 are distinct.  This large number of inequalities is not particularly enlightening to plot so we plot instead the simplest non-trivial case, $n=5$,
in Fig.~\ref{fig:simplespin}(b).

\begin{figure}
	\subfloat[]{{\includegraphics[height=5.3cm]{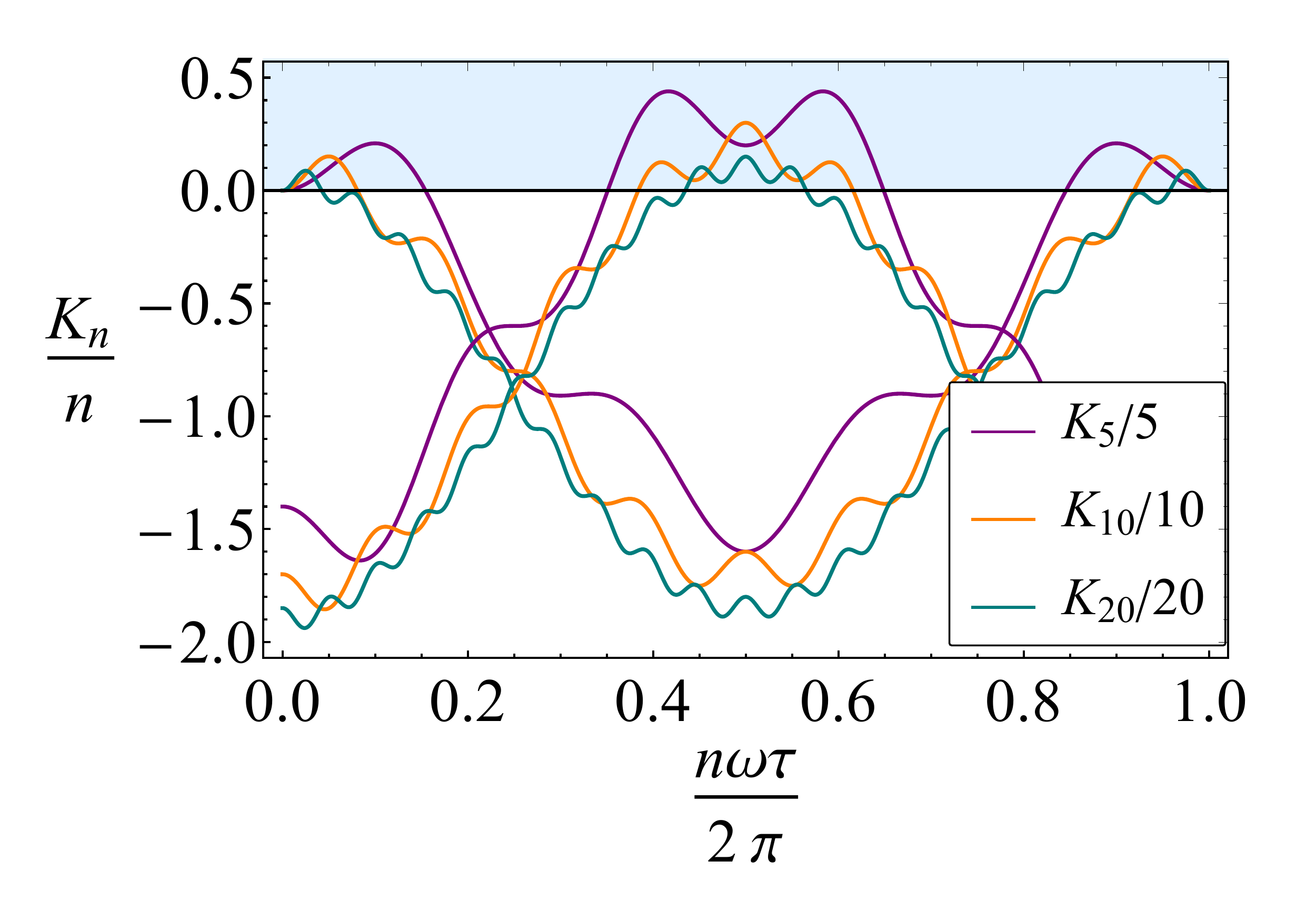}}}%
	\qquad
	\subfloat[]{{\includegraphics[height=5.3cm]{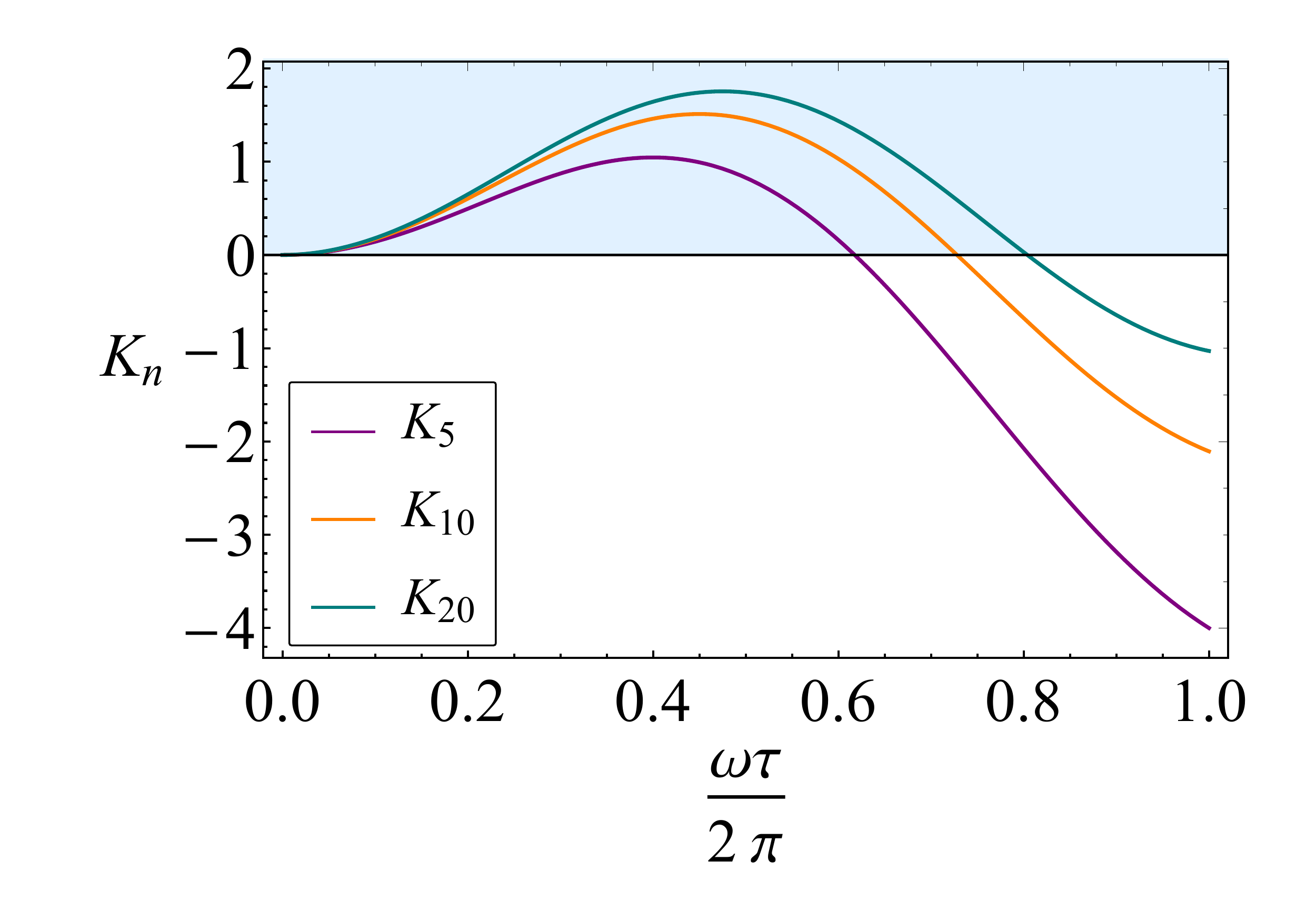}}}
	\caption{The violations of the ${\rm LG}_n$ inequalities within a simple quantum mechanical spin model.  We plot only the inequalities leading to the greatest violation, where violations are depicted by the shaded region.  In (a), measurements are performed in a way which extends the experimental time period, and in this regime the ${\rm LG}_n$ inequalities become easier to satisfy for higher $n$.  Note that although the violations appear smaller in magnitude, this is an effect of plotting the inequalities normalised by $n$.  In (b), more measurements are performed within the same experimental time period, and in this regime, the inequalities become harder to satisfy with increasing $n$.}%
	\label{fig:largen}%
\end{figure}

LG violations as a function of $n$ are shown in Fig.~\ref{fig:largen}.
The case in which increasing $n$ extends the time interval is plotted in Fig.~\ref{fig:largen}(a).  In the simple spin model, only two of the ${\rm LG}_n$ inequalities are violated for any $n$, both of which are plotted.  It can be seen that the region over which $K_n$ is violated decreases with $n$.  For the case in which the total time interval remains fixed as $n$ increases, only one of the ${\rm LG}_n$ inequalities is violated for any $n>4$, which is plotted in Fig.~\ref{fig:largen}(b).  Here the opposite effect is observed, where with increasing $n$, the region of violation {\it increases}

\begin{figure}
	\begin{center}
		\includegraphics[width=0.60\textwidth]{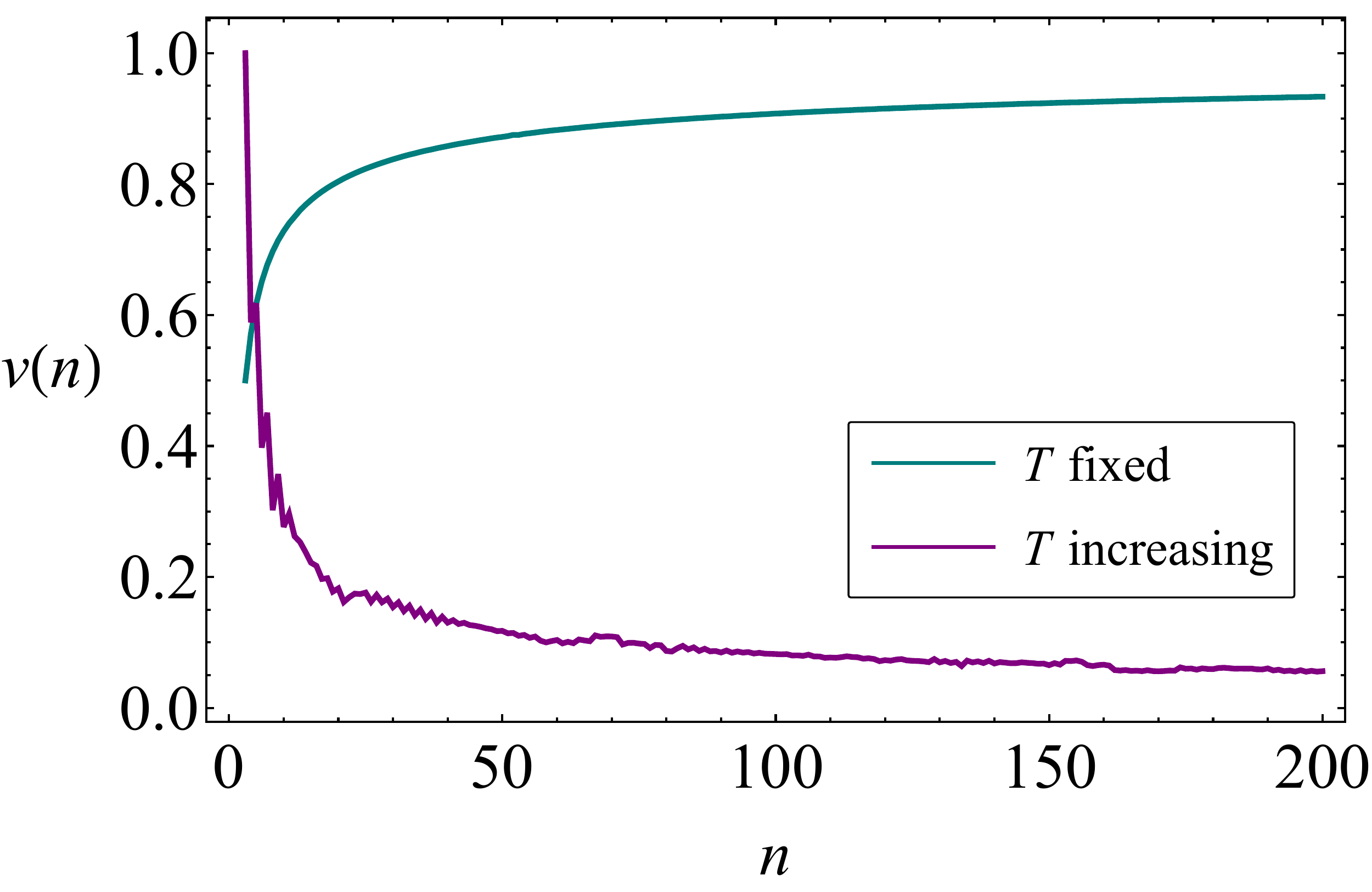}
	\end{center}
	\caption{
		\label{fig:nint}
		The fraction $\nu (n)$ of parameter space which leads to the violation of any of the ${\rm LG}_n$ inequalities  Eq.(\ref{lgansatz}) for the simple spin model.  Both the case where subsequent measurements subdivide the measurement window $[0,T]$ and where subsequent measurements increase the window $[0,T]$ are shown.  }
\end{figure}

We have also numerically calculated the fraction of parameter-space (i.e. range of times) over which {\it at least one} of the ${\rm LG}_n$ inequalities is violated, as a function of $n$.  This is plotted in Fig.~\ref{fig:nint}, for both cases, of either increasing or fixed time period with increasing $n$.  
This confirms the behaviour hinted at in Fig.~\ref{fig:largen}, that for the simple spin model, in the case of an increasing experimental time period the ${\rm LG}_n$ inequalities become easier to satisfy, whereas for the case of a fixed overall time period, the inequalities become harder to satisfy.

\section{Summary and Conclusions}

We have proved Fine's theorem for the case of $n$ measurements and in doing so established the general form
Eq.(\ref{lgansatz}) of a complete set of LG inequalities at $n$ times.
We considered generalizations of the usual LG approach to the situation in which all possible two-time correlators are measured and conjectured that necessary and sufficient conditions for the existence of an underlying probability consist of {the set of all three-time LG inequalities together with the $n$-gon inequalities on all $n$ correlators Eq.(\ref{neweq}).} We proved this conjecture for the cases $n=4$ and $n=5$ in the symmetric case.
This fills a gap in the previous work Ref.\cite{HalNIM} which explored a generalized LG approach in which higher order correlators are measured but omitted the extended two-time correlator case.
We explored both of the above sets of inequalities in the large $n$ limit and confirmed agreement with intuitive expectations.

The {$n$-gon inequalities} Eq.(\ref{neweq}) are clearly of interest to explore experimentally since they are qualitatively different to the usual LG inequalities. The $n$-time LG inequalities are of less interest experimentally, in comparison to experimental tests carried out already, since, as we saw here, they are logically equivalent to sets of three-time LG inequalities. 

However, the $n$-time LG inequalities could find another role since there are related problems that can be mapped onto the ${\rm LG}_n$ situation. LG inequalities usually concern a single dichotomic variable at each time but some recent works have explored situations involving variables taking three or more values (see for example, Ref.\cite{Ema}) which could easily be mapped onto the dichotomic case by adding extra times. Another possible application is in the contextuality by default approach to assessing the effects of signaling in LG inequalities
\cite{DK} in which the original three variables $Q_1$, $Q_2$, $Q_3$ in ${\rm LG}_3$ are adjoined with three more variables representing the values of the $Q_i$ in different contexts. This situation is equivalent to a $6$-time LG situation and can be analyzed using the inequalities derived here.
These possibilities will be studied in future papers.

\section{Acknowledgements}

We are grateful to Clive Emary, George Knee, Johannes Kofler, Raymond Laflamme, Shayan Majidy, Owen Maroney, Stephen Parrott and James Yearsley for many useful discussions and email exchanges about the Leggett-Garg inequalities. {We also thank an anonymous referee for useful comments}.

\bibliography{apssamp}

\begin{thebibliography}{10}

\bibitem{Bell} J.S.Bell, Physics (N.Y.) 1, 195 (1964), reprinted, along with most
of Bell's other key papers, in J.S.Bell, {\it Speakable and Unspeakable in Quantum
Mechanics} (Cambridge University Press, Cambridge, 1987).

\bibitem{CHSH} J.F.Clauser, M.A.Horne, A.Shimony and R.A.Holt, Phys.Rev.Lett. 23, 1306 (1982).
J.F.Clauser and A.Shimony, Rep. Prog. Phys. 41, 1881 (1978).


\bibitem{Per} A.Peres, {\it Quantum Theory: Concepts and Methods}
(Kluwer, Dordrecht, 1993).

\bibitem{LeGa} A.J.Leggett and A.Garg,
Phys. Rev. Lett. 54, 857 (1985).


\bibitem{L1} A. J. Leggett,
Foundations of Physics, 18, 939 (1988); 
Rep. Prog. Phys., 71, 022001 (2008).


\bibitem{ELN}  C. Emary, N. Lambert and F. Nori, Rep. Prog. Phys. 77, 016001 (2014)


\bibitem{AbBr} S.Abramsky and A.Brandenburger, New Journal of Physics, 13, 113035
(2011).

\bibitem{ConPhys}  R. W. Spekkens, 
Phys. Rev. A, 71, 052108 (2005).


\bibitem{ConPsy} 
I. Basieva, V. H. Cervantes, E. N. Dzhafarov, and K. Khrennikov, JEP:G (2019);
D. Aerts, L. Gabora, and S. Sozzo, Topics in Cognitive Science 5, 737 (2013);
M. Asano, T. Hashimoto, A. Y. Khrennikov, M. Ohya, and Y. Tanaka, Physica Scripta T163,
14006 (2014);
P. D. Bruza, K. Kitto, B. J. Ramm, and L. Sitbon, Journal of Mathematical Psychology 67,
26 (2015);
P. D. Bruza, Z. Wang, and J. R. Busemeyer, Trends in Cognitive Sciences 17, 383 (2015);
 J. M. Yearsley and J. J. Halliwell, arXiv:1905.12570.

\bibitem{DK} E. N. Dzhafarov and J. V. Kujala, arXiv 1407.2886v7 (2015).


\bibitem{Fine} A.Fine, J.Math.Phys. 23, 1306 (1982); Phys.Rev.Lett. 48, 291 (1982).


\bibitem{Bus} P.Busch, in {\it Non-locality and Modality} edited by. T. Placek, J. Butterfield, Springer-Verlag, NATO Science Series II. Mathematics, Physics and Chemistry
64, 175 (2002). (Also available as quant-ph/0110023).

\bibitem{SuZa} P.Suppes and M.Zanotti, Synthese 48, 191 (1981).

\bibitem{Pit} I.Pitowski, {\it Quantum Probability -- Quantum Logic}, Lecture Notes in Physics 321 (Springer-Verlag, Berlin, 1989).

\bibitem{GaMer} A.Garg and N.D.Mermin, Found.Phys. 14, 1 (1984).

\bibitem{ZuBr} M.Zukowski and C.Brukner, Phys. Rev. Lett. 88, 210410 (2002).

\bibitem{JHFine} J. J. Halliwell, Phys. Lett. A 378, 2945 (2014).


\bibitem{KoBr} J. Kofler and C. Brukner,
Phys. Rev. A 87, 052115 (2013).

\bibitem{Cle} L.Clemente and J.Kofler, Phys. Rev. A 91, 062103 (2015); Phys. Rev. Lett. 116, 150401 (2016).

\bibitem{HalQ} J.J.Halliwell, Phys. Rev. A 93, 022123 (2016).

\bibitem{HalLG4} J.J.Halliwell, Phys. Rev. A 96, 012121 (2017). A concise and updated version of this work is the e-print arXiv:1811.10408.



\bibitem{HaYe} J.J.Halliwell and J.M.Yearsley, Phys.Rev. A87, 022114 (2013).

\bibitem{Kly} D.N.Klyshko, Phys. Lett. A218, 119 (1996).


\bibitem{WLG} D. Saha, S. Mal, P. K. Panigrahi and D. Home, Phys. Rev. A, 91, 032117 (2015);
S. Kumari and A. K. Pan, 
Phys. Rev. A 96, 042107 (2017); S. Kumari and A. K. Pan, 
EPL 118, 50002 (2017).

\bibitem{Avis} D. Avis, P. Hayden and M. M. Wilde, Phys. Rev. A 82, 030102 (2010).

\bibitem{deg} C. Budroni and C. Emary, Phys. Rev. Lett. 113, 050401 (2014).

\bibitem{PQS} A. K. Pan, Md. Qutubuddin and S. Kumari, 
Phys. Rev. A 98, 062115 (2018).

\bibitem{Tsi} B. S. Tsirelson, 
Lett. Math. Phys. 4, 93 (1980).

\bibitem{HalNIM} J. J. Halliwell, Phys. Rev. A 99, 022119 (2019).



\bibitem{Bec} A. Bechtold, F. Li, K. Muller, T. Simmet, P.-L. Ardelt, J. J.
Finley, and N. A. Sinitsyn, Phys. Rev. Lett. 117, 027402
(2016).














\bibitem{deco1} M.M.Wilde and A.Mizel, Found.Phys 42, 256 (2012).

\bibitem{LGn2} M. Barbieri, 
Phys. Rev. A 80, 034102 (2009).


\bibitem{Ema} C.Emary, Phys. Rev. A 96, 042102 (2017).

\bibitem{CLT} P. Billingsley, {\it Probability and Measure} (3rd ed.), (John Wiley \& Sons, New York, 1995, p. 357
\end{thebibliography}

\end{document}